\documentclass[usegraphicx,usedcolumn,usenatbib]{mn2e}
\usepackage{graphicx}
\usepackage{txfonts}
\usepackage{rotating}
\usepackage{accents}

\newcommand{\apjl}{ApJL}

\newcommand{\mnras}{MNRAS}

\newcommand{\Msun}{M$_{\sun}$}

\newcommand{\HI}{H\,{\sc i}}

\newcommand{\Moy}{M$_{\sun}$\,yr$^{-1}$}

\newcommand{\MD}{$M_{\textrm{\scriptsize D}}$}

\newcommand{\Mvir}{$M_{\textrm{\scriptsize vir}}$}

\bibpunct[,]{(}{)}{;}{a}{,}{,}

\begin{document}

\title[Gas infall and the evolution of disc galaxies]
{The role of gas infall in the evolution of disc galaxies}

\author[Moll\'{a} et al. ]
{Mercedes ~Moll\'{a}$^{1}$\thanks{E-mail:mercedes.molla@ciemat.es},
{\'A}ngeles I. D{\'\i}az$^{2,3}$, Brad  K. Gibson$^{4}$, Oscar Cavichia$^{5}$,
\newauthor
{\'A}ngel-R. L{\'o}pez-S{\'a}nchez$^{6,7}$\\
$^{1}$ Departamento de Investigaci\'{o}n B\'{a}sica, CIEMAT,
Avda. Complutense 40. E-28040 Madrid. (Spain)\\ 
$^{2}$ Universidad Aut\'{o}noma de Madrid, 28049, Madrid, Spain \\
$^{3}$ Astro-UAM, Unidad Asociada CSIC, Universidad Aut{\'o}noma de Madrid, 28049, Madrid, Spain\\
$^{4}$ E.A. Milne Centre for Astrophysics, University of Hull, Hull, HU6~7RX, United Kingdom\\
$^{5}$ Instituto de F\'{i}sica e Qu\'{i}mica, Universidade Federal de Itajub\'{a}, Av. BPS, 1303, 37500-903, Itajub\'{a}-MG, Brazil\\
$^{6}$Australian Astronomical Observatory, PO Box 915, North Ryde, NSW 1670, Australia\\ 
$^{7}$Department of Physics and Astronomy, Macquarie University, NSW 2109, Australia}

\date{\today}

\pagerange{\pageref{firstpage}--\pageref{lastpage}} \pubyear{2016}
\maketitle
\label{firstpage}

\begin{abstract}
Spiral galaxies are thought to acquire their gas through a protracted 
infall phase resulting in the inside-out growth of their associated 
discs.  For field spirals, this infall occurs in the lower density 
environments of the cosmic web. The overall infall rate, as well as the 
galactocentric radius at which this infall is incorporated into the 
star-forming disc, plays a pivotal role in shaping the characteristics 
observed today.  Indeed, characterising the functional form of this 
spatio-temporal infall in-situ is exceedingly difficult, and one is 
forced to constrain these forms using the present day state of galaxies with 
model or simulation predictions.  We present the infall rates 
used as input to a grid of chemical evolution models spanning the mass
spectrum of discs observed today.  We provide a systematic
comparison with alternate analytical infall schemes in the literature,
including a first comparison with cosmological simulations.
Identifying the degeneracies associated with the adopted infall rate 
prescriptions in galaxy models is an important step in the development 
of a consistent picture of disc galaxy formation and evolution.
\end{abstract}


\section{Introduction}
Numerical chemical evolution models (CEMs) are one of the most flexible 
(and long-standing) tools for interpreting the distribution of metals in 
both the gas and stellar phases of galaxies.  The power of CEMs lies in 
the rapid and efficient coupling of star formation and feedback 
prescriptions, with the stellar nucleosynthesis, initial mass function 
formalisms, and treatments of gas infall/outflow.  While lacking a 
self-consistent (hydro-)dynamical treatment, the ability to explore these 
parameter spaces on computing timescales of minutes rather than months, ensures 
CEMs maintain a prominent role in astrophysics today.

The initial motivation for the development of CEMs was the
identification of what is now known as the {\it G-dwarf Problem}
\citep{vdb62,sch63,lyn75}; specifically, there is an apparent paucity
of metal-poor stars in the solar neighbourhood, relative to the number
predicted to exist should the region behave as a `closed-box',
i.e. one in which gas neither enters nor departs.  It was recognised
that a viable solution to the G-dwarf Problem lay in the relaxation of
this closed-box assumption, via the inclusion of a gas infall
prescription \citep{lar72,edm90}. The infall of metal-poor gas can
then dilute the existing elemental abundances, whilst simultaneously
increasing the early star formation rate, thus producing a stellar
metallicity distribution shifted to moderately higher
metallicities. Moreover the star formation sustained by metal-poor gas
accretion self-regulates to produce a constant gas-phase metallicity
close to the stellar yield (i.e., close to the solar
metallicity). The G-dwarf problem also appears in external galaxies, as
M\,~31 \citep{wor96}. Building on this framework, most classical models
of disc formation assume a protogalaxy or dark matter halo which acts
as the the source of the infalling gas
\citep{lf83,gus83,lf85,mat89,pcb98} without which, in addition, 
radial abundance gradients are increasingly difficult to 
recover \footnote{Some more recent models include two or even three 
infall phases, each
corresponding in turn to the formation of the halo, followed by that
of the thick and thin discs \citep{chia01,chia03,fen06,mic13}.}.

Extensions of this classical framework include those employing a
multiphase representation for the interstellar medium
\citep{fer92,fer94,mol96}.  In \citet{md05}, we calculated a generic
grid of theoretical CEMs, defined in terms of their rotation velocity
using the universal rotation curve of \citet{pss96}.  In that work we
assumed that the infall rate, or its inverse, the collapse time-scale,
$\tau_{\textrm{\scriptsize c}}$, for each galaxy, depends on the total
mass of each theoretical galaxy, with the low mass galaxies forming on
a longer time scale than the massive ones, according to the expression
$\tau_{\textrm{\scriptsize c}} \propto M^{-1/9}$ \citep{gal84}.  Such
a mass-dependency mimics the downsizing phenomenon now associated with
galaxy formation \citep{heav04,gon08}.

These sorts of adopted timescale relationships are, however, weakly 
constrained in the sense that they were only implemented to ensure 
present-day abundance patterns and gas fractions were recovered.  We now 
possess much more information pertaining to the manner by which gas 
moves from the cosmic web, through halos, and onto discs, and so more 
realistic prescriptions should be pursued.

From the observational point of view, the observations of the atomic
hydrogen line \HI\ at 21~cm in nearby galaxies revealed the existence
of extensive halos containing gas up to 15\,kpc above the plane of the
discs \citep[e.g.][]{fra02,bar05,boo08,hea11,gen13}.  This gas could
be deposited on the discs, forming at present the reservoirs from
which stars form in the outer parts of galaxies.  In fact, it seems be
rotating slower than in discs and moving slowly toward inwards
\citep{oos04}.  However, most of observational evidences of gas
accretion are indirect, since this gas is tenuous. Only the high
velocity clouds (HVC) are well studied \citep[e.g.][]{wak99}, their
existence being in agreement with theoretical expectations.
\citet{ric06} analyzed the HVCs in the local intergalactic medium
suggesting, in agreement with previous findings from \citet{bli99},
that these HVC's are the building blocks of galaxies, and situated
within the halo at galactocentric distances less than $\sim$40~kpc,
consistent with the predictions of cosmological hydrodynamical
simulations \citep{con06}.  These clouds are comprised of
low-metallicity gas \citep[e.g.][]{gib01}, consistent with the
material being the fuel out of which the disc forms.  It should also
be noted that there now exists recent observations which suggest the
infall of low-metallicity \HI\ gas in dwarf galaxies triggering star
formation therein.  The most prominent example is NGC~5253, but
additional examples are described by \citet[][ and references
therein]{arls10,arls12}.\footnote{Similarly, it has been suggested
that the high N/O ratio found in some of these galaxies might be also
related to infall of metal-poor gas. See the excellent review by
\citet{jorge14} for details.}
 
\citet{oos04} estimated that, if bumps of \HI\ gas have a total mass
of order 10$^{8-9}\,\mbox{M}_{\sun}$ and are accreted in
10$^{8-9}$\,yr, the typical accretion rate would be $\sim$1\,\Moy.  On
the other hand, \citet{san08}, reviewing all data referring this
subject of gas around galaxies and its possible movement towards them,
infer a mean visible accretion rate of cold gas of at least 0.2\,\Moy,
which should be considered a lower limit of the infall rate; such a
value poses a problem when it compared with the star formation rate
(SFR), since it is roughly one order of magnitude smaller than
necessary to sustain the observed SFR \citep[see][ ,and references
therein]{jorge14}. The \citet{san08} value is, however, calculated
using only the cold gas, neglecting the likely more dominant
background reservoir of ionised gas.  \citet{leh11} measuring the mass
of this phase, estimating the infall rate increases to $\sim
0.8$\,\Moy.  Later, \citet{ric12} gives an infall rate of
0.5--0.7\,\Moy\ as an estimate for the Milky Way Galaxy (MWG) and the
Andromeda galaxy (M\,31) within a radius of 50\,kpc, claiming that `in
MWG and other nearby galaxies the infall of neutral gas may be
observed directly by \HI\ 21~cm observations of extra-planar gas
clouds that move through the halos'. Very recently, \citet{fer16}
report the {\sc HI } 21-cm detection in emission at a redshift
$z=0.375$ with the COSMOS {\sc HI} Large Extragalactic Survey
(CHILES).  Following these data, the diffuse gas is
M(HI)=2$\times$10$^{10}$\,\Msun, the molecular one is
M(H$_{2}$)=1.8--9.9$\times$10$^{10}$\,\Msun, and the stellar mass
M$_{*}$=8.7$\times$10$^{10}$\,\Msun.  This implies that the disc of
this galaxy accreted [13--22]$\times$10$^{10}$\,\Msun\, in
$\sim$8\,Gyr, which will produce an averaged value for the infall rate
of 17--27\,\Msun\,yr$^{-1}$.  Although these numbers are highly
speculative, given the lack of any more firm estimates, we will employ
them in this work.

Additional guidance regarding infall rates can be provided by tracing
the spatial and temporal infall of gas onto discs, via the use of
cosmological simulations. This cosmological gas supply has a strong
dependence on redshift and halo mass. However, the interplay between
the circumgalactic gas components is not well known and the gas
physics in a turbulent multiphase medium is non-trivial to capture in
(relatively) low-resolution hydrodynamical simulations. The
consequence is that there remain only select examples in the
literature which reproduce successfully and simultaneously (both in
terms of size and relative proportions) the characteristics of
late-type discs and their spheroids \citep[e.g.][]{brad,vog13,scha15}.
Bearing in mind this cautionary statement, it is interesting to note
that these simulations suggest that most of the baryons in galaxies
are accreted diffusely, with roughly 3/4 due to smooth accretion, and
1/4 from mergers.

Recently, \citet{bro14} have generated a suite of cosmological
simulations which reproduce the gross characteristics of the Local
Group.  Analysing these simulations, they find the relationship
between the stellar mass and the halo mass (their Fig.~2 and Eq.~2),
valid for a stellar mass range [$10^{7}-10^{8}$] M$_{\sun}$.  In the
last decade, several techniques have been developed to obtain such a
relation between the dynamical mass in the proto-halos and the
baryonic mass in the discs, although usually this last one is
associated to the stellar mass.  One of these statistical approaches
connecting the CDM halos with their galaxies is the sub-halo abundance
matching technique. With that, the total stellar to halo mass relation
(SHMR) is obtained \citep{shan06,guo10,beh10,beh13,rod12,rod15}.
Other formalisms use the halo occupation distribution (HOD), which
specifies the probability that a halo of mass $M$ has a given number
of galaxies with a certain mass $M_{*}$ (or luminosity, colour or
type). As a result, the SHMR is also estimated
\citep{mos10,leau10,yang12}. A summary of these results can be found
in Fig.\,5 of \citet{beh13}. This relation constrains the possible
accretion of gas from the halo to the discs.

 One of the questions that arises when cosmological simulations and
data are compared is, such as \cite{kor16} states, that there is a
collision between the cuspy central density seen in cosmological
simulations and the observational evidence that galaxies have flat
cores. This tension there exists from some years ago and it is still
present. The use of a Navarro, Frank \& White (NFW) profile, with its
$\rho \propto r^{-1}$ producing cusps at small radii, comes from the
era in which cosmological simulations were undertaken primarily only
with dark matter, but it continues be widely used. Over the past years
the data for rotation curves have improved immensely, as well as the
mass modeling, showing that most of dwarf disc galaxies have cored
halos. Although it seemed less clear for giant spirals, \citet{don09}
analyzed rotation curves (RCs) for a sample of 1000 galaxies finding a
good fit of a core-halo profile to the data, better than the one for
the NFW. More recently, \citet{nesti13} have carefully analyzed the
available data for the Milky Way Galaxy, fitting both dark matter
Burkert and NFW profiles. They find that the cored profile produces
the best result, and is therefore the preferred one, claiming that
this is in agreement with similar fits obtained for other external
disc galaxies and in agreement with the mass model underlying the
Universal Rotation Curve (URC). \citet{ogi14a} and \citet{ogi14b} say
that this discrepancy between observations and simulations may be due
to dynamical processes that transform a cuspy into a cored model,
probably by the effect of the feedback that modifies the star
formation process at small scale. In fact, most recent cosmological
simulations \citep[see][]{brad} which include this feedback in the
star formation prescriptions, find that this transformation occurs
when there is violent feedback from rapid star formation in the inner
regions of disc galaxies. In this work we use the \citet[][
hereinafter SAL07]{sal07} expressions, who use the URC formalism
assuming that halo distributions follow a Burkert core isothermal
profile.

In the present work we compute the infall rate for a set of
theoretical galaxies with total dynamical masses in the range \Mvir
$\sim[5\times 10^{10} -10^{13}]$ \Msun.  Following the prescriptions
of SAL07, we derive the rotation curves for each halo and disc, and
their corresponding radial mass distributions.  By imposing that gas
from the halo falls onto the discs at a rate such that after a Hubble
time the systems end with masses as observed in nature, we obtain the
infall rate for each galaxy and for each radial region therein.  We
analyse the infall rate resulting from these prescriptions, comparing
with the results from assumptions of previous CEMs, those inferred
from cosmological simulations, and those from extant empirical data
concerning mass accretion. We pay special attention to the redshift
evolution of this infall in galaxies of different dynamical masses,
and analyse its radial dependency within individual galaxies. We
verify that the final halo-disc mass relation follows the
prescriptions given by the authors cited above.  The chemical
evolution is beyond the scope of this work; here, we focus
specifically on the manner by which gas reaches the disc.  The impact
on star formation and metal enrichment is the focus of the next phase
of our collaboration (Moll\'{a} et al., in preparation).

We describe the framework of our models in Section~\ref{model}. The
results are outlined in Section~\ref{results}, sub-dividing the study
of the dependence of the infall rate on the galactocentric radius into
Sub-section ~\ref{dep-r}, the dependence on mass of the whole galaxies
into Sub-section ~\ref{dep-m}, and the resulting growth of the spiral
discs into Sub-section~\ref{mdisc}. These results and their
implications are discussed in \S~\ref{dis}. Our conclusions are
summarised in Section~\ref{conclusions}.

\section{Model Framework}
\label{model}

For our calculations we use the SAL07 equations, the details for which
are outlined in that work.  These authors combine kinematic data of
the inner regions of galaxies with global observational properties to
obtain the URC of disc galaxies and the
corresponding mass distributions.  For that they use a universal halo
density profile following \citet{bur95}, while the disc is described
by the classical Freeman surface density law. With both components,
they compute the two rotation curves that contribute to the total.
Using data from rotation curves for around 1000 disc galaxies, and
fitting them to the above described theoretical rotation curve, they
estimated the URC and correlations between the observational
properties and the parameters defining those RCs. These are the
expressions given and used here. The SAL07 work is the continuation of
that of \citet{pss96}, with the difference being that now the results
are given directly as functions of \Mvir, instead of any other
observational quantity.

We start by assuming an initial mass of gas in a spherical region or
protogalaxy.  The total dynamical masses for our theoretical galaxies
are taken in the range \Mvir$ = [5\times 10^{10}-10^{13}]$\,\Msun,
with values starting at $\log{M_{\textrm{\scriptsize vir}}}=10.75$,
increasing in mass in steps of $\Delta\log{M_{\textrm{\scriptsize
vir}}}=0.15$, resulting in a total set of 16 models.

For each \Mvir, the virial radius, $R_{\textrm{\scriptsize vir}}$, is
computed.  The latter is defined as being the radius corresponding to
the transition between the virialised matter of a given halo and that
of the infalling material; formally, this corresponds to:
\mbox{$R_{\textrm{\scriptsize vir}}=259\,(M_{\textrm{\scriptsize
vir}}/10^{12}{\rm M_{\sun}})^{1/3}$\,kpc}.

To link the virial mass with the baryonic mass in the disc, \MD, the
relationship from \citet[][ hereinafter SHAN06]{shan06} is used:
\begin{equation} 
M_{\textrm{\scriptsize star}}\,[{\rm M_{\sun}}]=2.3\,\times 10^{10} \frac{\left(M_{\textrm{\scriptsize vir}}/3.10^{11}\,
{\rm \mbox{M}_{\sun}}\right)^{3.1}}{1+\left(M_{\textrm{\scriptsize vir}}/3.10^{11}\,{\rm \mbox{M}_{\sun}}\right)^{2.2}}
\label{shankar06}
\end{equation}
\noindent
In order to obtain the final baryonic disc mass, \MD, it is necessary
to include the mass tied up in the gas phase. From SHAN06, the atomic
gas mass is related with the $B$-band luminosity\footnote{Similar
relationships can be seen between the atomic gas mass and the $R$-band
luminosity - e.g., Fig.~8(c) of \citet{Brook12b}.}, and therefore with
the stellar mass, as:
$\log{M_{\textrm{\scriptsize HI}}}=2.42+0.675\,\log{M_{\textrm{\scriptsize star}}}$; 
through these relationships, we compute 
$M_{\textrm{\scriptsize D}}=M_{\textrm{\scriptsize star}}+ 1.34\times M_{\textrm{\scriptsize HI}}$
\footnote{In what follows, we ignore the contribution of the
molecular gas component to $M_{\textrm{\scriptsize{D}}}$.}.

Both the virial and disc masses possess an intrinsic radial
distribution, and therefore, following again SAL07, we use the
rotation curves to compute their respective radial dependencies, as:

\begin{equation} V^{2}(R)=V_{\textrm{\scriptsize
H}}^{2}(R)+V_{\textrm{\scriptsize D}}^{2}(R)
\end{equation}
\begin{eqnarray} 
V_{\textrm{\scriptsize H}}^{2}(R)\,[{\rm km\,s^{-1}}] & = & 6.4G\frac{\rho_{0}R_{0}^{3}}{R} \bigg\{ \ln \bigg(1+\frac{R}{R_{0}} \bigg) \nonumber \\ 
& & -{\rm atan}\frac{R}{R_0} + \frac{1}{2} \ln \bigg( 1 + \frac{R^2}{R_{0}^{2}} \bigg) \bigg\},
\end{eqnarray} 
where $\rho_{0}$ is the central density of the halo:
\begin{equation} 
\log{\rho_{0}\,[{\rm g\,cm^{-3}}]}=-23.773-0.547\log{\frac{M_{\textrm{\scriptsize vir}}}{10^{11}\,{\rm M_{\sun}}}},
\end{equation} 
and $R_{0}$ is the core radius for the Burkert profile,
taken from \citet{yeg12} (where Eq. 10 from SAL07 was updated),
and given by:
\begin{equation}
 \log{R_{0}\,[{\rm kpc}]} =0.71+0.547\log{\frac{M_{\textrm{\scriptsize vir}}}{10^{11}\,{\rm
M_{\sun}}}}
\end{equation} 

and
\begin{eqnarray} 
V_{\textrm{\scriptsize D}}^{2}(R)\,[{\rm km\,s^{-1}}]&=&\frac{1}{2}\frac{GM_{\textrm{\scriptsize D}}}{R_{\textrm{\scriptsize D}}}
(3.2x)^{2}(I_{0}K_{0}-I_{1}K_{1}),
\end{eqnarray} 
where $x=R/R_{\textrm{\scriptsize opt}}$, the optical radius is defined as: 
$R_{\textrm{\scriptsize opt}}=3.2R_{\textrm{\scriptsize D}}$; and $I_{n}$ and $K_{n}$ are the
modified Bessel functions computed at 1.6$x$.  
The scale-length of the disc, $R_{\textrm{\scriptsize D}}$, is given by the expression:
\begin{eqnarray} 
\log{R_{\textrm{\scriptsize D}}\,[{\rm kpc}]}=0.633+0.379\log{\frac{M_{\textrm{\scriptsize D}}}{10^{11}\,{\rm{M}_{\sun}}}} \nonumber\\+0.069\left(\log{\frac{M_{\textrm{\scriptsize D}}}{10^{11}\,{\rm{M}_{\sun}}}}\right)^{2}
\end{eqnarray} 

We define a characteristic radius for each model as
$R_{\textrm{\scriptsize c}}=R_{\textrm{\scriptsize opt}}/2$.  This
radius, as $R_{\textrm{\scriptsize opt}}$ and $R_{\textrm{\scriptsize
D}}$, are, however, used only for normalization purposes, and for this
work are not related to the surface brightness profile.

Having computed the components of rotation velocity, the radial mass
distributions within the halo and disc components, for each value of
\Mvir, are given by:
\begin{eqnarray} 
M_{\textrm{\scriptsize H}}(<R) &= &2.32\,10^{5}R\,V_{\textrm{\scriptsize H}}(R)^{2}\\
M_{\textrm{\scriptsize D}}(<R) &= &2.32\,10^{5}R\,V_{\textrm{\scriptsize D}}(R)^{2}
\end{eqnarray}

To these two components we add a bulge component.  To compute this
term, we use the fact that correlations between disc and bulge
structural parameters exist \citep{bal07,gan09}. From these
correlations, we obtain the following expressions for the central
velocity dispersion, $\sigma_{0}$, and the effective radius of the
bulge, $R_{\textrm{\scriptsize e}}$:
\begin{eqnarray}
\sigma_{0}\,[{\rm km\,s^{-1}}] & = &  105\, R_{\textrm{\scriptsize D}}^{0.54},\\
R_{\textrm{\scriptsize e}}\,[{\rm kpc}] & = &  0.32\, R_{\textrm{\scriptsize D}}-0.045
\end{eqnarray}

Then:
\begin{equation}
M_{\textrm{\scriptsize bulge}}[{\rm M_{\sun}}]=2.32\,10^{5}\sigma_{0}^{2}\,R_{\textrm{\scriptsize e}},
\end{equation}
and 
\begin{equation}
M_{\textrm{\scriptsize bul}}(R)[{\rm M_{\sun}}]=2.32\,10^{5}[\sigma_{0}\,\textrm{e}^{-R/R_{\textrm{\scriptsize e}}}]^{2}\,R.
\end{equation}
In the central region ($R=0$), we have also added the mass
corresponding to the supermassive black hole, $M_{\textrm{\scriptsize
BH}}$, following the classical expression: $\log{M_{\textrm{\scriptsize
BH}}}=\beta \log(\sigma/220)+\alpha$, with $\alpha= 8$ and $\beta=4$
\citep{chan13}.

Finally, we then have:
\begin{equation}
M_{\textrm{\scriptsize tot}}(R)=M_{\textrm{\scriptsize D}}(R)+M_{\textrm{\scriptsize H}}(R)+M_{\textrm{\scriptsize bul}}(R).
\end{equation}

\begin{table*}
\scriptsize
\caption{Characteristics of the theoretical galaxies modeled in this work.}
\begin{tabular}{rrrrrrrrrrrrr}
\hline
N  & Name  &   $M_{\textrm{\scriptsize vir}}$  & $M_{\textrm{\scriptsize D}}$  &  $M_{\textrm{\scriptsize bulge}}$ &  $ R_{\textrm{\scriptsize vir}}$ & $R_{\textrm{\scriptsize D}}$ & $R_{\textrm{\scriptsize opt}}$ & $R_{\textrm{\scriptsize c}}$ & $\tau_{\textrm{\scriptsize c}}$ & $V_{\mbox{rot}}$  & $\sigma_{0}$ & $R_{\textrm{\scriptsize e}}$ \\
 &           & [$10^{10}$\, \Msun] & [$10^{10}$\,\Msun] & [$10^{10}$\,\Msun] & [kpc] & [kpc] & [kpc]  & [kpc]  &   [Gyr]   & [km s$^{-1}$]  & [km s$^{-1}$] & [kpc] \\
\hline
 1  &   10.75   &     5.62 &     0.023 &    0.064  &     99.228   &    1.298  &     4.154  &     2.077 &    106.769 &     43.667 &    122.046  &     0.382   \\
 2  &   10.90   &     7.94 &     0.056 &    0.070  &    111.335   &    1.347  &     4.312  &     2.156 &     48.801 &     47.231 &    124.703  &     0.398   \\
 3  &   11.05   &    11.22 &     0.14  &    0.087  &    124.920   &    1.469  &     4.701  &     2.351 &     25.758 &     55.764 &    131.078  &     0.438   \\
 4  &   11.20   &    15.85 &     0.33  &    0.122  &    140.163   &    1.674  &     5.357  &     2.679 &     16.057 &     70.210 &    141.338  &     0.506   \\
 5  &   11.35   &    22.39 &     0.75  &    0.184  &    157.265   &    1.967  &     6.295  &     3.148 &     11.982 &     89.870 &    155.119  &     0.602   \\
 6  &   11.50   &    31.62 &     1.51  &    0.288  &    176.455   &    2.334  &     7.470  &     3.735 &      9.901 &    112.342 &    171.199  &     0.723   \\
 7  &   11.65   &    44.67 &     2.67  &    0.454  &    197.985   &    2.744  &     8.781  &     4.390 &      8.773 &    134.799 &    187.931  &     0.858   \\
 8  &   11.80   &    63.10 &     4.24  &    0.680  &    222.143   &    3.172  &    10.150  &     5.075 &      8.054 &    155.754 &    204.317  &     0.999   \\
 9  &   11.95   &    89.13 &     6.25  &    1.014  &    249.249   &    3.618  &    11.577  &     5.788 &      7.549 &    175.306 &    220.413  &     1.145   \\
10  &   12.10   &    125.9 &     8.82  &    1.495  &    279.662   &    4.098  &    13.114  &     6.557 &      7.158 &    194.300 &    236.844  &     1.303   \\
11  &   12.25   &    177.8 &    12.19  &    2.192  &    313.786   &    4.636  &    14.835  &     7.417 &      6.843 &    213.516 &    254.300  &     1.480   \\
12  &   12.40   &    251.2 &    16.67  &    3.340  &    352.073   &    5.255  &    16.815  &     8.408 &      6.586 &    233.462 &    273.360  &     1.684   \\
13  &   12.55   &    354.8 &    22.70  &    5.250  &    395.033   &    5.980  &    19.135  &     9.567 &      6.378 &    254.425 &    294.508  &     1.922   \\
14  &   12.70   &    501.2 &    30.84  &    8.579  &    443.234   &    6.837  &    21.880  &    10.940 &      6.213 &    276.557 &    318.181  &     2.205   \\
15  &   12.85   &    660.7 &    39.40  &    13.18  &    485.996   &    7.641  &    24.451  &    12.225 &      6.110 &    295.146 &    339.232  &     2.469   \\
16  &   13.00   &    1000  &    56.87  &    27.08  &    557.999   &    9.087  &    29.079  &    14.539 &      5.999 &    324.533 &    374.896  &     2.945   \\
\hline
\label{input}
\end{tabular}
\end{table*}
\normalsize

The 16 calculated models would produce discs with baryonic masses $
M_{\textrm{\scriptsize D}}$ in the range $\sim [2.3 \times
10^{8}-5.5\times10^{11}]$\,\Msun.  The observed galactic masses for
which SAL07 estimated the relations used here, lie between 10$^{9}$
and $2\times10^{11}$~\Msun.  Therefore, the first two models have
lower masses (2.3 and 5.6$\times$$10^{8}$\,\Msun, respectively)
outside this range. On the massive end of the spectrum, we have
$\sim$3 models with masses above this limit, which could therefore be
considered more appropriately as spheroids or lenticulars.  As such,
some caution should be applied when considering the mass extrema of
our models.

Table~\ref{input} summarises the characteristics of the radial mass
distributions, obtained from the above expressions, that define our
theoretical protogalaxies.  For each model with number N (column~1),
we give the logarithm of the virial mass, which is used to name each
model, and the value of this mass, $M_{\textrm{\scriptsize vir}}$ in
columns 2 and 3; the mass the disc would have at the end of the
evolution, $M_{\textrm{\scriptsize D}}$, is listed in in column 4; the
mass of the bulge, $M_{\textrm{\scriptsize bul}}$, in column 5; the
virial radius, $R_{\textrm{\scriptsize vir}}$, in column 6; the disc
scale length, $R_{\textrm{\scriptsize D}}$, the optical radius,
$R_{\textrm{\scriptsize opt}}$, and the characteristic radius,
$R_{\textrm{\scriptsize c}}$, are in columns 7 to 9. The value of the
collapse timescale at the characteristic radius,
$\tau_{\textrm{\scriptsize c}}$, (see next subsection) is in column
10; the maximum rotation velocity of the disc for each model,
$V_{\textrm{\scriptsize rot}}$, in column 11; the velocity dispersion,
$\sigma_{0}$, at the center of the bulge in column 12; and the
effective radius of the bulge, $R_{\textrm{\scriptsize e}}$, in column
13.

\begin{figure}
\includegraphics[width=0.495\textwidth,angle=0]{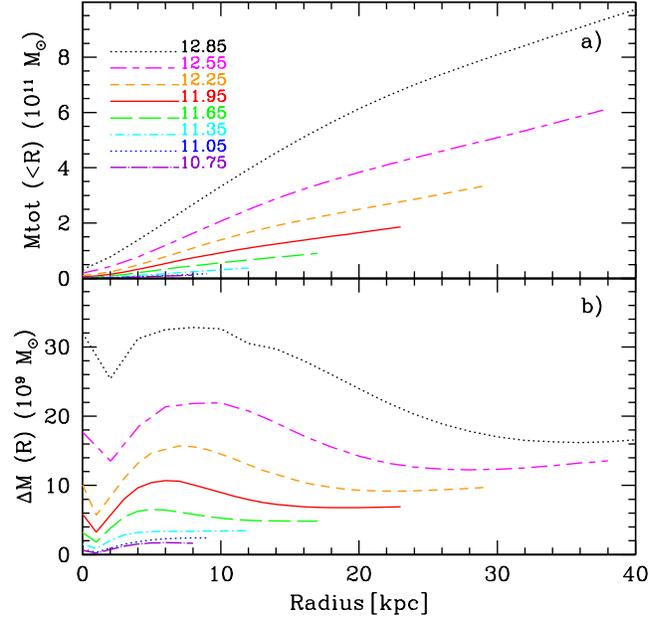}
\caption{a) Radial distributions of total mass $M_{\textrm{\scriptsize
tot}}(<R) $ for different values of virial mass \Mvir, given by their
logarithm as labelled.  b) Radial distribution of the mass $\Delta
M_{\textrm{\scriptsize tot}}(R)$ within each radial region of modelled
galaxies.}
\label{masas}
\end{figure}

The total mass radial distributions $M(R)$ are represented in
Fig.~\ref{masas} for selected values of $M_{\textrm{\scriptsize
vir}}$ labelled with their logarithm values in panel a). In panel b)
we show the mass included in each radial region, which would be a
cylindrical region above and below the corresponding annulus in the
disc or equatorial plane in which the gas will fall; that is, $\Delta
M_{\textrm{\scriptsize tot}}(R)=M_{\textrm{\scriptsize
tot}}(<R)-M_{\textrm{\scriptsize tot}}(<R-1)$.

\section{Infall rates}
\label{results}

\subsection{The infall rate radial distributions: prescriptions for the collapse timescale}
\label{dep-r}

\begin{figure}
\includegraphics[width=0.495\textwidth,angle=0]{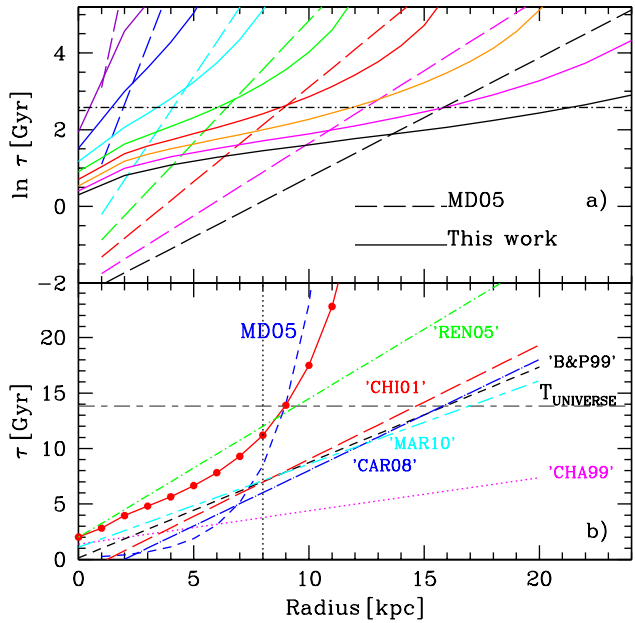}
\caption{a) Dependence of the collapse timescale $\tau$ with 
galactocentric radius R.  Each solid line represents a given radial
mass distribution for a given \Mvir, represented with the same colours as in
Fig.~\ref{masas}.  The long dashed lines represent the radial
distributions of the collapse timescales for similar models from the
MD05 grid.  b) Comparison of the radial dependence of the collapse
time scale of the most similar to MWG model, corresponding to $\rm
N=9$, shown by the solid red line and dots for this work, compared
with the old models shown by the short-dashed blue line, labelled as
MD05, and with other radial functions from other authors,
\citet{cha99,boi99,chia01,ren05,car08}, and \citet{mar10}, labelled as
CHA99, B\&P99, CHIA01, REN05, CAR08, and MAR10, respectively.}
\label{tcol_r}
\end{figure}

We assume that the total mass is initially in a spherical region in
the gas phase, from which it `falls' from the halo onto the equatorial
plane.  As a consequence of this infall, the disc is formed at a
characteristic timescale called the `collapse timescale',
$\tau_{coll}(R)$, defined as the time necessary for the disc mass
$M_{\textrm{\scriptsize D}}(R)$ to reach the actual value at the
present time through the prescribed infall formalism.  In other words,
the halo mass decreases in a Hubble time by the exact quantity that
goes into the disc via infall, integrated over each radial bin.  We
have already calculated the initial or total mass in the halo, $\Delta
M(R)$, (see Fig.1), in each radial region, and we also know what the
corresponding $\Delta M_{\textrm{\scriptsize D}}(R)$ must be. We may,
therefore, calculate the collapse time for each radial region as:

\begin{equation}
\tau(R)=-\frac{13.2}{\ln{\left(1-\frac{\Delta M_{\textrm{\scriptsize D}}(R)}{\Delta M_{\textrm{\scriptsize tot}}(R)}\right)}}\,[\mbox{Gyr}]
\end{equation}

With the knowledge that the collapse timescale depends on the dynamics
of the gas, and that spiral discs possess clear radial density
profiles, it should not be surprising to realise that $\tau$ will have
a radial dependence.  Some others have already included this radial
dependence for the infall rate in their models
\citep{lf85,mat89,pcb98,boi00,ren05,fen06}, by assuming different
expressions. In fact, such a radial dependence is inherent to
classical `inside-out' disc formation scenarios, and is essential for
obtaining the observed density profiles and radial abundance
gradients. Since the mass density seems to be an exponential, we had
also assumed an exponential expression for the collapse timescale in
MD05, with a steep dependence with galactocentric radius; conversely,
other chemical evolution models employed more conservative linear
dependencies on galaxy radius.
\begin{figure}
\includegraphics[width=0.38\textwidth,angle=-90]{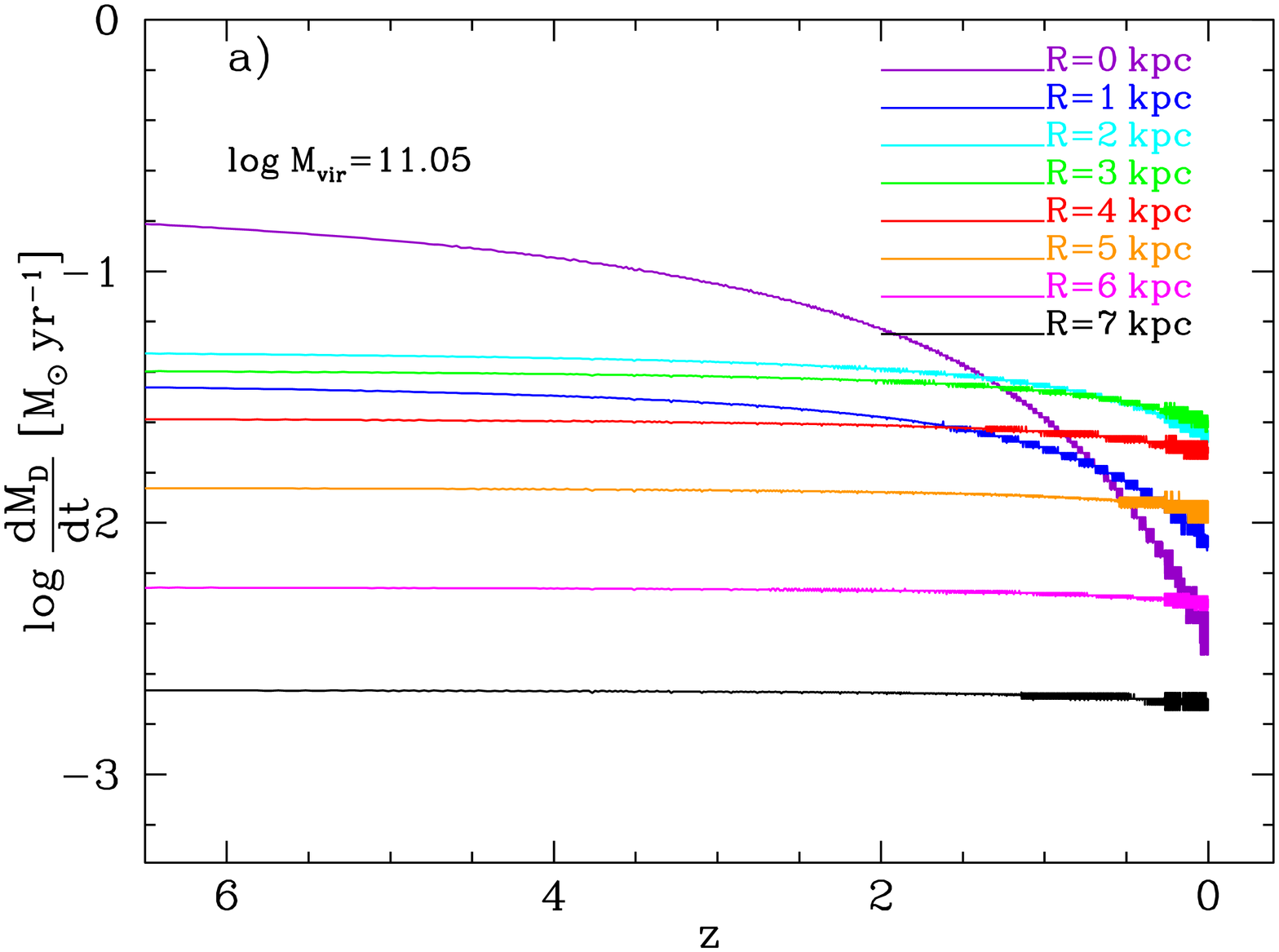}
\includegraphics[width=0.38\textwidth,angle=-90]{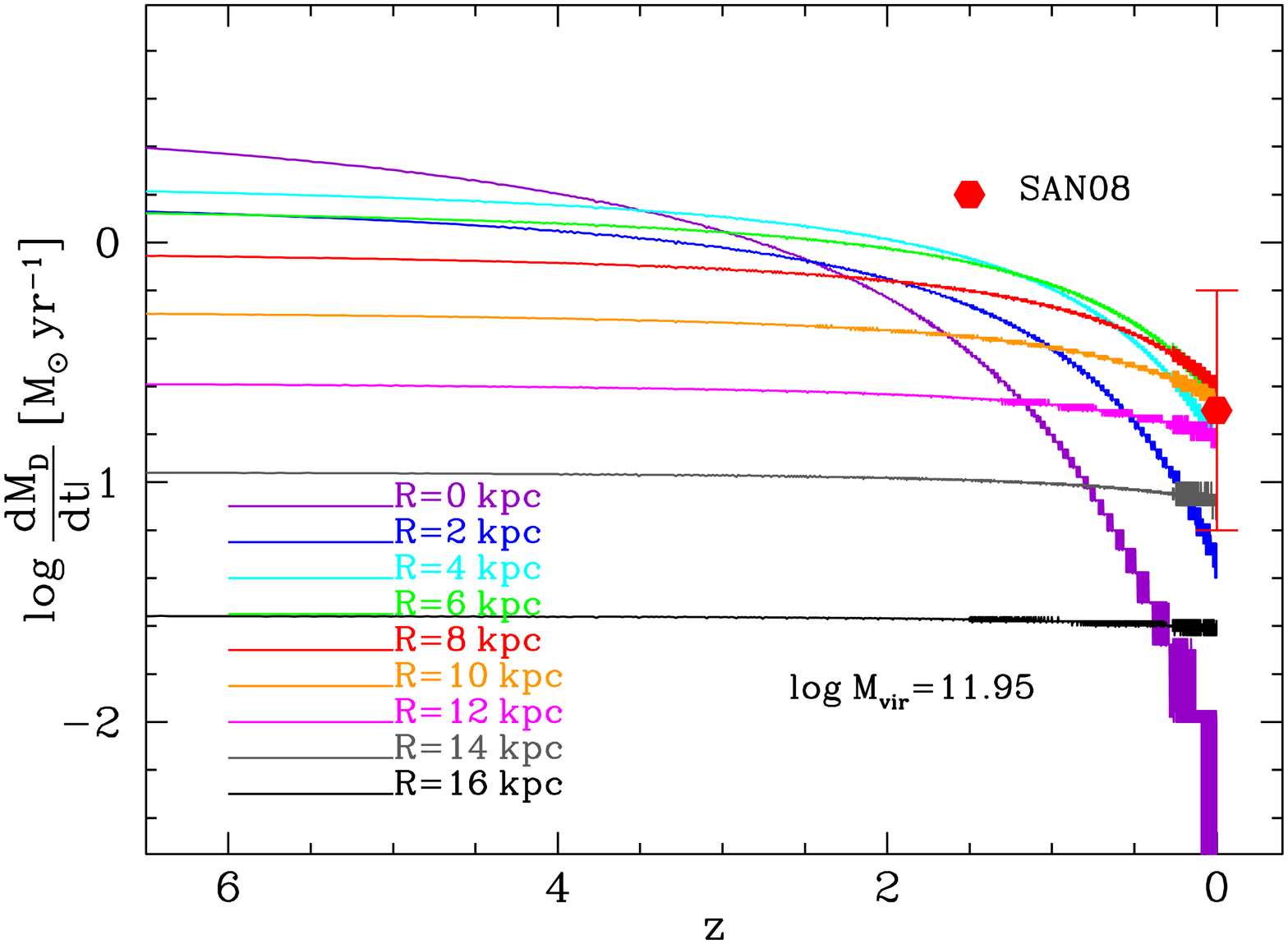}
\includegraphics[width=0.38\textwidth,angle=-90]{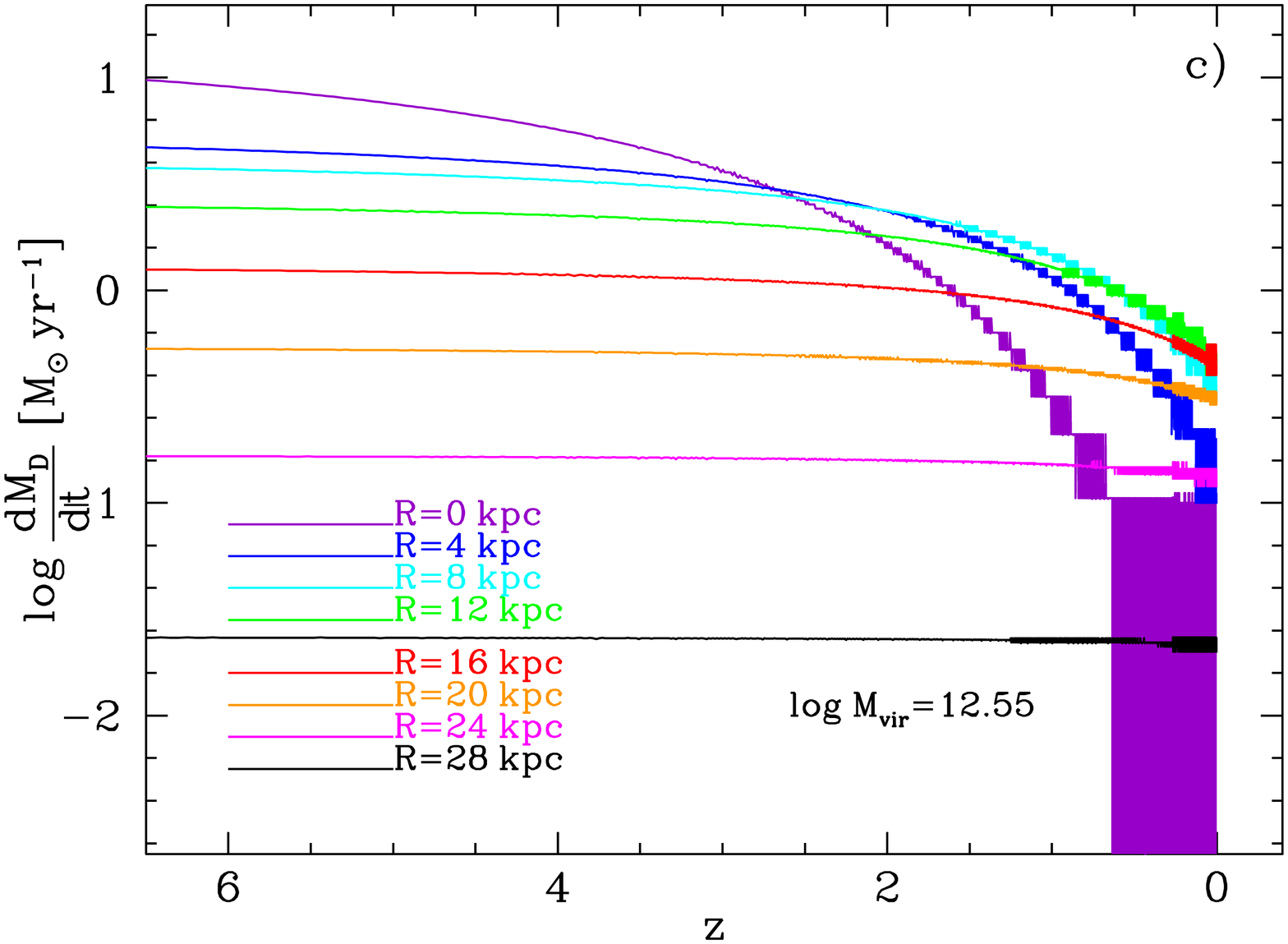}
\caption{The infall rate in logarithmic scale as a function of the
redshift $z$ for several radial regions, as labelled, in: a) a
low-mass galaxy ($\log{M_{\textrm{\scriptsize vir}}}=11.05$); b) a
MWG-type galaxy ($\log{M_{\textrm{\scriptsize vir}}}=11.95$); and c) a
massive galaxy ($\log{M_{\textrm{\scriptsize vir}}}=12.55$). The red
point with error bars is the observational estimate for the MWG
\citep{san08}.}
\label{infall-r}
\end{figure}

The resulting radial distribution of $\tau(R)$ for this new grid is
shown in Fig.~\ref{tcol_r}. In panel a), these values are shown, in
logarithmic scale, for the same models as in Fig.~\ref{masas},
represented with the same colours and line coding. Solid lines are the
results of this work, while long-dashed lines are the collapse
time-scale used in MD05, for the closest model in rotation velocity.
Functionally, the new (non-linear) collapse timescales are very
different from the older ones (upper panel).  The most central regions
($<$2~kpc) show a linear behaviour, linking the bulge with the more
'curved' expression corresponding to the disc.  One can see that the
inner regions now have now longer timescales compared with the MD05
prescriptions, while in the outer discs, the values are more
comparable (although typically slightly shorter when compared with the
older models).

In the panel b) we show the curve corresponding to the MWG-like model,
($\rm N=9$) for this work, compared with our previous MD05 model for
the MWG (model NDIS=28) and other (all linear) forms drawn from the
literature (and labelled accordingly).  One can see that the red line
(our model, here) shows longer collapse timescales at a given radius
than the other models, except for the inner disc of the \citet{ren05}
model (although the difference between these two is minimal at these
galactocentric radii).
 
As a consequence of this varying collapse timescale with radius, a
different infall rate is produced in each radial region, building the
disc in an inside-out fashion over a Hubble time.  In
Fig.~\ref{infall-r}, we show the evolution with redshift of the
predicted infall rate, $dM_{\textrm{\scriptsize D}}/dt$, for regions
located at different galactocentric radii in: a) a low mass galaxy, b)
a galaxy representing the MWG, and c) a massive galaxy. We see that
the very central regions -- purple in a) or purple and blue lines in
b) and c) -- where the bulge is located, show a strongly variable
infall rate compared with the disc regions, where the curves are
essentially flat until $z=2$, and then slightly decreasing to $z=0$.
The black lines for $R=7$, 16, and 28\,kpc are below the observational
data. This is interesting, mainly if we take into account that the
optical radius in MWG is $\sim 13-14$\,kpc; that is, the infall rate
seems to define well the size of the disc.  An additional
characteristic highlighted by this figure is that even with different
absolute values, the infall rate behavior is very similar in all
radial regions of all discs, and clearly different to that of the
bulge regions.

\begin{figure}
\includegraphics[width=0.48\textwidth,angle=0]{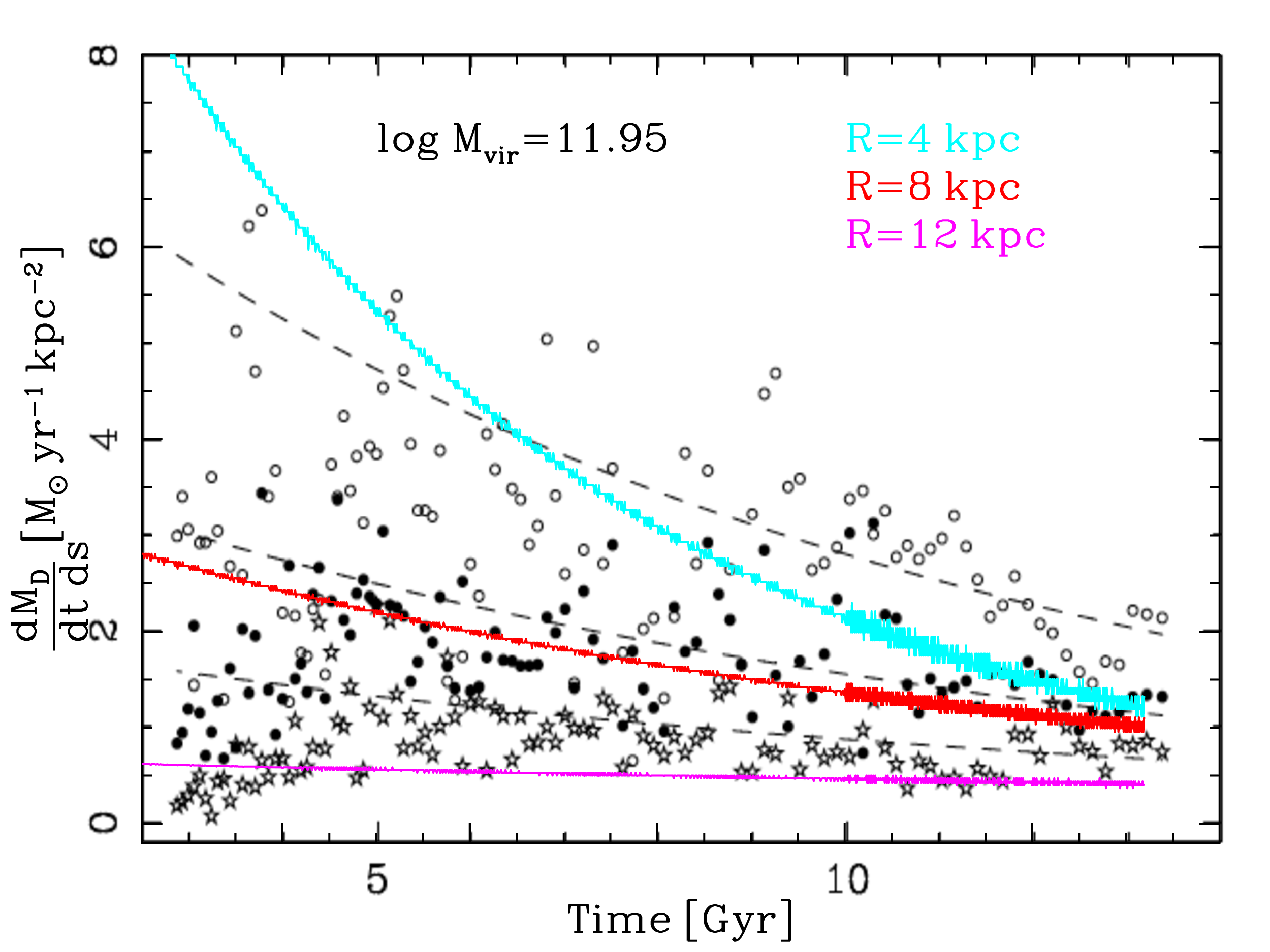}
\caption{The time evolution of the infall rate density as (dM/dt\,dS) in
\Msun\,$yr^{-1}\,kpc^{-2}$ units, for three different radial regions,
as labelled, compared with results from \citet{courty10} for
galactocentric radii 4.5, 8.5, and 12.5\,kpc, as open dots, filled dots,
and stars, respectively. The dashed line are similar radial region
results drawn from the chemical evolution model of \citet{chia97}. }
\label{new}
\end{figure}

The comparison of the detailed infall rates from our models with
cosmological simulations is not straightforward; the public
availability of simulation suites such as Illustris \citep{nel15b}
will aid in the future, but even that amazing data-set has spatial
resolution limitations which makes tracing sub-100pc disk 'impact'
positions for infalling gas very difficult.  And being a moving mesh,
rather than particle-based, tracing the temporal history of said
infalling gas is non-trivial.  We have undertaken a cursory initial
comparison though with the cosmological mesh simulation of
\citet{courty10}, as a demonstrator for what will be a more ambitious
comparison in the future.  For this particular simulation, the authors
use multi-resolved, large-scale structure, N-body/hydrodynamical
simulations, whose initial conditions are re-centered on a Milky Way
sized halo with M$_{dyn}=7.2\times 10^{11}$~M$_\odot$. The
simulations include, in addition to gravitation and gas dynamics, star
formation and its associated thermal and kinetic feedback from
supernovae. They identified a large reservoir of gas in the halo
fueling the disk within the virial radius, and quantified the gas
accretion rate by computing the gas flowing through spherical surfaces
or slabs located at different galactocentric distances (or distances
above the mid-plane, in the case of slabs). In Fig.~\ref{new} we have
represented our infall rates fluxes (that is, as infall rate surface
densities), corresponding to log$M_{vir}=11.95$ (M$_{vir}=8\times
10^{11}$~M$_\odot$), as a function of time, for three radial regions
located at inner, $\sim$solar, and outer regions, plotted with
different colours, as labelled.  We have included here the results
from \citet{courty10}, (their Fig.3, right panel), estimated (only
account for the hot gas) for three galactocentric radii of 4.5, 8.5,
and 12.5\,kpc, shown by open dots, filled dots, and stars,
respectively, and also the results for \citet{chia97}, drawn as dashed
lines for similar radial regions. We see that our results reproduce
the same behavior found by cosmological simulations: a clear evolution
of the infall rates decreasing with time, and also with radius at a
given time. This is expected for an inside-out formation process for
the galactic disk.

\subsection{The dependence on the dynamical mass: the halo-disc ratio}
\label{dep-m}

We defined in Section~2, a characteristic radius,
$R_{\textrm{\scriptsize c}}$, that we use in comparing radial mass
distributions of different sizes. We have calculated the collapse
timescale for this particular radius, interpolating in the radial
distributions of $\tau(R)$, thus finding the characteristic collapse
timescale, $\tau_{\textrm{\scriptsize c}}$, which is in column 10 of
Table 1, for each galaxy.
\begin{figure}
\includegraphics[width=0.38\textwidth,angle=-90]{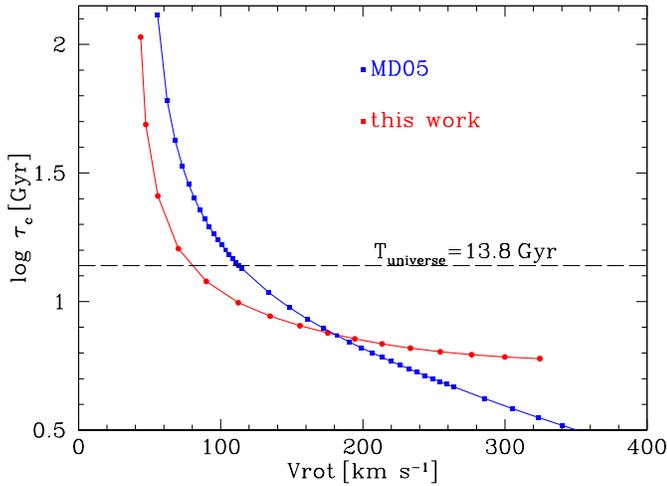}
\caption{Relation of the characteristic collapse time scale
$\tau_{\textrm{\scriptsize c}}$ with the maximum rotation velocity of
galaxies.  The dashed (grey) line shows the time corresponding to the
age of the Universe.}
\label{tcol-m}
\end{figure}

We plot in Fig.~\ref{tcol-m} this $\tau_{\textrm{\scriptsize c}}$ as a
function of the maximum rotation velocity of the disc. The results of
our work here (red) are compared with those employed in MD05 (blue),
computed, as said, by the expression from \citet{gal84}.  Although a
similar dependence on the total mass still appears, with shorter
$\tau_{\textrm{\scriptsize c}}$ for massive galaxies and longer times
for the low-mass galaxies, differences among both grids appear in both
ends of the maximum rotation velocity range. For massive galaxies
(essentially more massive than MWG), $\tau_{\textrm{\scriptsize c}}$
is now longer than in the older MD05 models, while it is shorter for
the lower-mass galaxies.

\begin{figure}
\includegraphics[width=0.495\textwidth,angle=0]{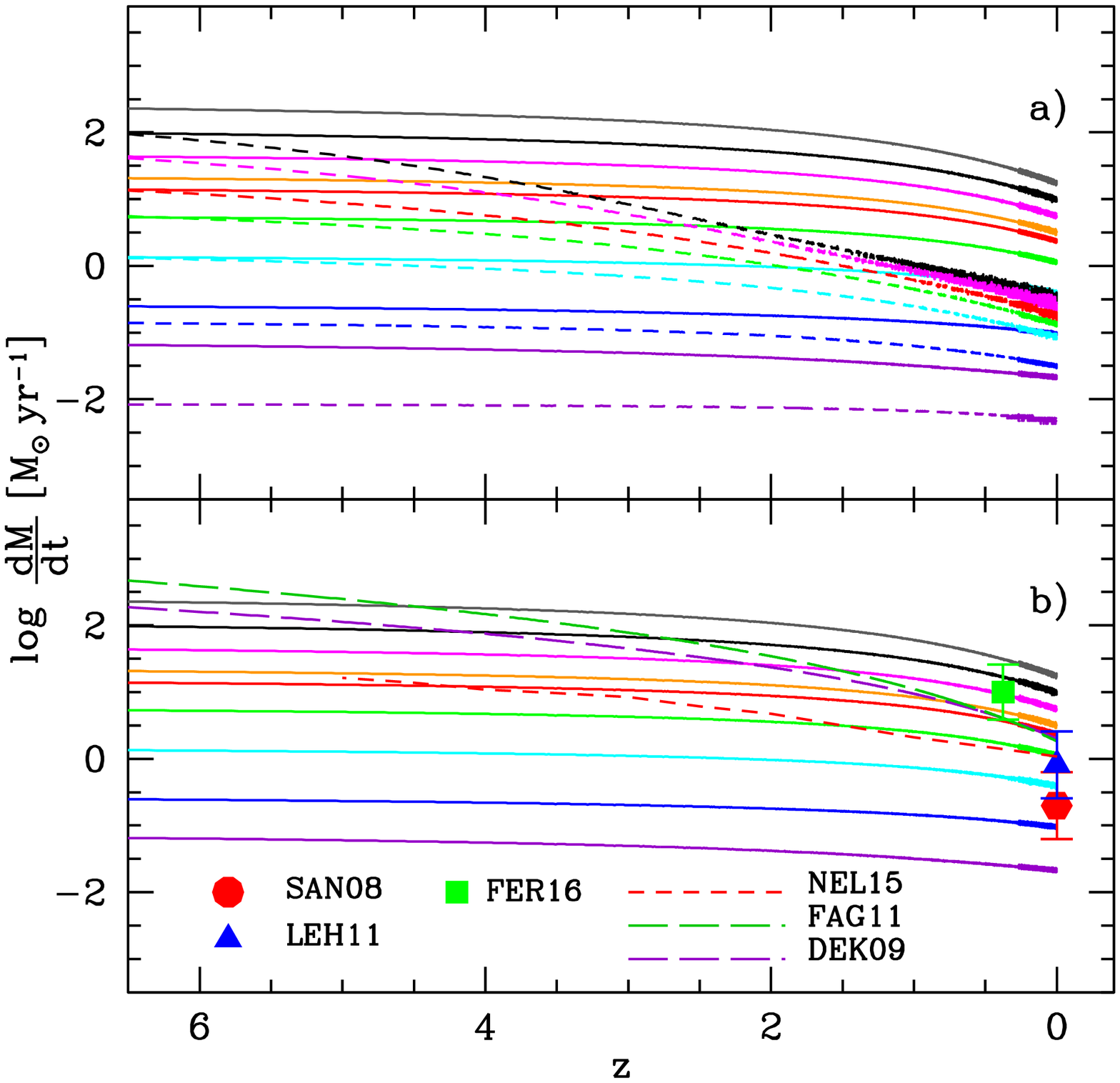}
\caption{The infall rate as a function of the redshift $z$ for galaxies
of several virial mass, with the same coding as Fig.1. a): comparison with
MD05 infall rates shown by short-dashed lines. b): comparison with
cosmological simulations, as long-dashed lines, from \citet[][
DEK09]{dek09} and \citet[][ FG11]{fau11}, and as short-dashed line
from \citet[][ NEL15]{nel15b}, as labelled. The solid red
pentagon is the estimated value from \citet{san08} while the blue
triangle is given by \citet{leh11}. The green square is obtained from 
\citet{fer16} for the highest redshift galaxy observed in {\sc HI}.}
\label{infall-m}
\end{figure}

We now integrate the infall across all radial zones to derive the
infall rate for the whole disc for each galaxy. This way we may
compare the total infall rate for different virial mass galaxies. Such
as we see in Fig.~\ref{infall-m}, where this total infall rate for the
same examples of Fig.~\ref{masas} are represented as solid lines, the
gas falls from the halo to the disc with a different rate depending on
the total mass, as expected from Fig.~\ref{tcol-m}. In panel a) we
compare these with the models from MD05 (short-dashed lines), where it
is clear that the infall rate now is more constant in time and,
therefore, maintains higher values (than previously) for the more
massive galaxies, at the present time.

Panel b) of Fig.~\ref{infall-m} compares our results with the
cosmological simulations presented by \citet{dek09} and
\citet{fau11}. Both sets of simulations are clearly more variable in
time, in better agreement with our old models, but with higher
absolute values. It is necessary though to note that these particular
simulations lead to primarily massive spheroidal systems (rather than
late-type discs).  The cosmological simulations from Illustris
\citep{nel15a} produce discs more similar to those observed in nature.
In \citet{nel15b}, the team analyse how galaxies acquire their gas in
simulations, with and without feedback. They found that the time taken
for the gas to cross the virial radius increases by a factor of
$\sim$2-3 in the presence of feedback, but is independent of the halo
mass (being in the range 10$^{10}$ to 10$^{12}$ \Msun), as we also
found. Using their Fig.\,8, we plot in Fig.~\ref{infall-m} the
redshift evolution of the non-feedback simulation. We see that these
results show a much lower infall rate than the old spherical galaxy
simulations, and more in agreement with our (slightly) flatter models.
It is necessary to be clear, however, that this line corresponds to
the accretion corresponding to a virial mass log\, \Mvir=11.30; that
is, it may be compared with our cyan line, which lies below. It
implies that this model will create a more massive disc than ours' for
the same virial mass. We have also shown the standard value of the
infall rate obtained for the HVCs in the MWG \citep{san08}. This
value, which reproduces well the infall prediction for the solar
region, is, however, lower than expected when the total infall for the
closest-to-MWG simulated galaxy (red or green line) is considered. In
that case, the value of \citet{leh11} is, however, well reproduced.
The value at $z=0.375$ is an estimate obtained from the recent work by
\citet{fer16} who detect {\sc HI} at the highest redshift, to
date. They give the masses for the diffuse (and molecular gas), and
use the stellar mass from Spitzer IRAC data:
M(HI)=2$\times$10$^{10}$\,\Msun, the molecular one is
M(H$_{2}$)=1.8--9.9$\times$10$^{10}$\,Msun, and the stellar mass
M$_{*}=8.7\times 10^{10}$\,\Msun.  Using these numbers, a total disc
mass is estimated in the range [13.4--21.5]\,$\times$10$^{10}$\Msun.
If the disc accreted this mass in $\sim$8\,Gyr (the evolutionary time
from $z$$\sim$7 to today), the averaged value for the infall rate
would be $\sim$17-27\,\Msun\,yr$^{-1}$.  Taking into account that this
is an averaged value across the full redshift range, and that the
infall rate probably was greater at higher redshift, we reduce the
value by a factor of two, that is $\log{(dM/dt)}=1.00$. We note that
taking into account the total mass of the disc, this galaxy would lie
between our models 11 and 13, corresponding to $\log{M_vir}=12.25$ and
12.55 (represented by magenta and orange lines in the accompanying
figure).  We see that these recent observational data compares very
well with our model predictions.

\begin{figure}
\includegraphics[width=0.495\textwidth,angle=0]{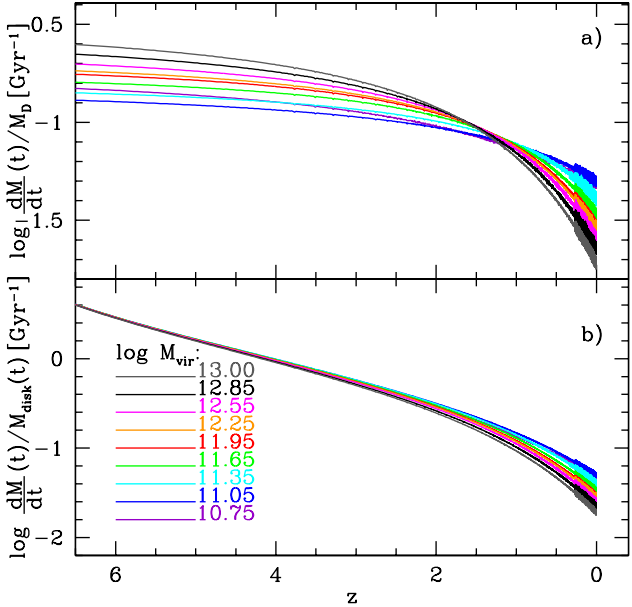}
\caption{The infall rate normalised to a) the final mass of the disc
$\frac{dM}{dt}/M_{\textrm{\scriptsize D}}$, and b) to the disc mass at
each time $M_{\textrm{\scriptsize D}}(t)$, both in logarithmic scale,
as a function of the redshift $z$ for the same galaxies as shown in
Fig.~\ref{infall-m}.}
\label{infall-nor}
\end{figure}

The infall rate behaviour is very similar for all galaxy masses, with
differences mainly in the absolute value.  As such, we have normalised the
infall rate to the final mass of the disc, $M_{\textrm{\scriptsize
D}}$ and show the results in Fig.~\ref{infall-nor}. In a) we normalise
to the final disc mass $M_{\textrm{\scriptsize D}}$ while in panel b)
we normalise to the disc mass at each time $M_{\textrm{\scriptsize
D}}(t)$. We see that in panel a) all infall rates coincide to the same
value $\frac{dM}{dt}/M_{\textrm{\scriptsize D}}\sim 0.1\,Gyr^{-1}$, at
$z=1.3$, while in panel b) it is clear that the last normalised infall
rate is practically the same for $z> 2.5$. This means that at high
redshift, discs grow in the same proportion for all virial masses and
it is only recently that differences appear.

\begin{figure}
\includegraphics[width=0.495\textwidth,angle=0]{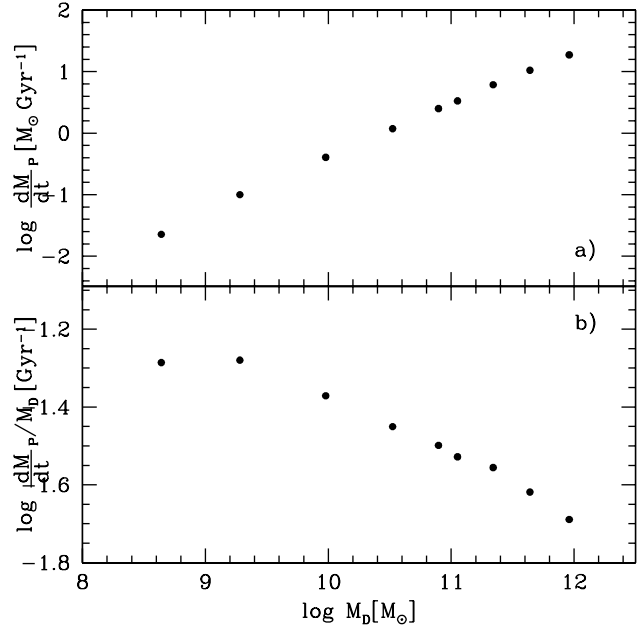}
\caption{a) The infall rate at the present time ($z=0$),
$({\frac{dM}{dt}})_{P}$ and b) as normalised to the final mass of the
disc at the present time $({\frac{dM}{dt}})_{P}/M_{\textrm{\scriptsize
D}}$ as a function of disc mass $M_{\textrm{\scriptsize D}}$.}
\label{infall-masa}
\end{figure}

The infall rate at the present time shows a correlation with the
virial mass of the dark halo or with the mass in the disc
$M_{\textrm{\scriptsize D}}$, as we show in Fig.~\ref{infall-masa}.
In panel a), we see that the infall rate depends on the mass,
increasing with $M_{\textrm{\scriptsize D}}$. However, in panel b),
where the normalised infall rate is represented, the contrary occurs:
low mass galaxies are now suffering a higher infall rate in proportion
to the total mass of their discs, while the massive discs have now a
very low rate (almost an order of magnitude lower). This is again in
agreement with the scenario where low mass galaxies form their discs
well after the most massive ones (the latter of which create their
discs rapidly).
 
\subsection{The growth of spiral discs}
\label{mdisc}

As a consequence of this gas infall scenario, the disc is formed in an
inside-out fashion.  The proportion of the final mass in the disc
compared with the total dynamical mass of the galaxy is dependent on
this total mass, in a consistent way to our inputs for calculating
$M_{\textrm{\scriptsize D}}$ (included as parameter in the rotation
curves).

\begin{figure}
\includegraphics[width=0.48\textwidth,angle=0]{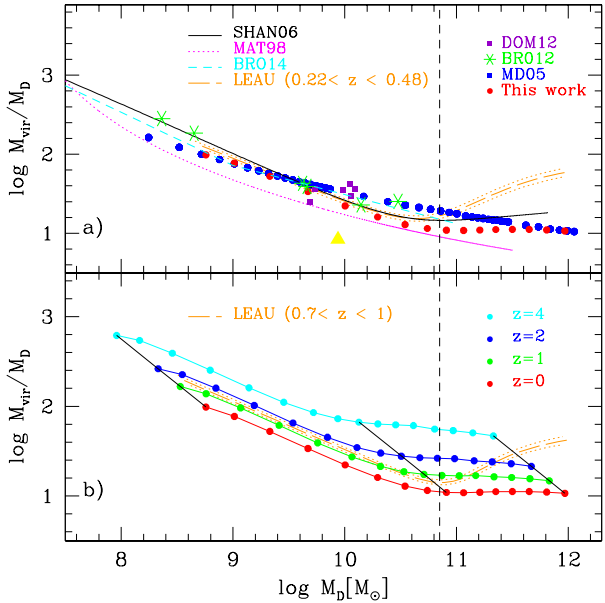}
\caption{a) The ratio $\frac{M_{var}}{M_{\textrm{\scriptsize D}}}$ at
$z=0$ as a function of the mass in the disc $M_{\textrm{\scriptsize
D}}$. Our models, calculated to follow the SHAN06 prescription (black
line), are the red dots, while the blue dots are the MD05 results.
The magenta, cyan, and black lines are the prescriptions obtained by
\citet{mateo98}, labelled as MAT98, and from BRO14 and SHAN06 from
observations of the Local Group of galaxies for the first two, and
from halo distribution data, for the latter. The orange line shows
results from LEAU for COSMOS data for the lowest redshift range.
Cosmological simulations are drawn with purple squares and green stars
from DOM12 and BRO12, respectively; b) Our results for the ratio
$\frac{M_{var}}{M_{\textrm{\scriptsize D}}}$ at $z=0$, 1, 2, and 4,
represented with different colours, as labelled, as a function of the
mass in the disc $M_{\textrm{\scriptsize D}}$ at each time, compared
with results from LEAU for $z \sim 0.7-1.0$ (orange dashed line).
Black lines join the evolutionary tracks for individual galaxies with
$\log{M_{\textrm{\scriptsize vir}}}=10.75$, 11.95, and 13.00.}
\label{md_mt}
\end{figure}

Panel a) of Fig.~\ref{md_mt} shows the fraction of the virial to the
disc mass, $\frac{M_{\textrm{\scriptsize vir}}}{M_{\textrm{\scriptsize
D}}}$, resulting from the applied collapse timescale prescriptions of
our new grid (red points), compared with our previous results from
\cite{md05} (blue points).  We also show the relationship from
\citet{mateo98}, and from \citet[][ hereinafter BRO14]{bro14}, both
obtained for galaxies of the Local Group.  Our older results from MD95
had a similar slope to the one from \cite{mateo98}, but the absolute
value given by the latter is somewhat smaller, which is due to the
$\rm M_{*}/L$-ratio employed to transform the observations
(luminosities) to stellar masses.  The cosmological simulations begin
to be able to predict discs of the appropriate size. We have drawn in
the same Fig. ~\ref{md_mt}, the results obtained by \citet[][
hereinafter DOM12]{mariola} and by \citet[][ hereinafter BRO12]{brad}.
We also plot the results obtained by \citet[][ hereinafter
LEAU]{leau10}, who studied the stellar-to-halo mass relation using
COSMOS data, for the lowest redshift they give (as labelled in the
figure).  Our new models are calculated to predict the line from
SHAN06 and are close to the one from LEAU, and, in excellent agreement
with DOM12, BRO12, and BRO14.  In the cases of LEAU and SHAN06, lines
show an increase for high disc masses which, obviously, is not
apparent in our old models, since we had assumed a continuous
dependence of the collapse time scale with the dynamical mass. This
increase is, however, rather modest for the SHAN06 line and the models
follow this trend.  In panel b) of the same figure we show our results
for redshifts $z=0$, 1, 2, and 4, with different colours, as labelled.
The relation looks similar for $z \le 2$, however, when we see the
track for a particular galaxy, shown by black lines for three values
of $\log{M_{\textrm{\scriptsize vir}}}=10.75$, 11.95, and 13.00, the
evolution is clear, with decreasing ratios $M_{\textrm{\scriptsize
vir}}/M_{\textrm{\scriptsize D}}$ for increasing
$M_{\textrm{\scriptsize D}}$. Results for $z=1$ are close to the LEAU
data for the range $0.7 < z < 1.0$.

\begin{figure}
\includegraphics[width=0.45\textwidth,angle=0]{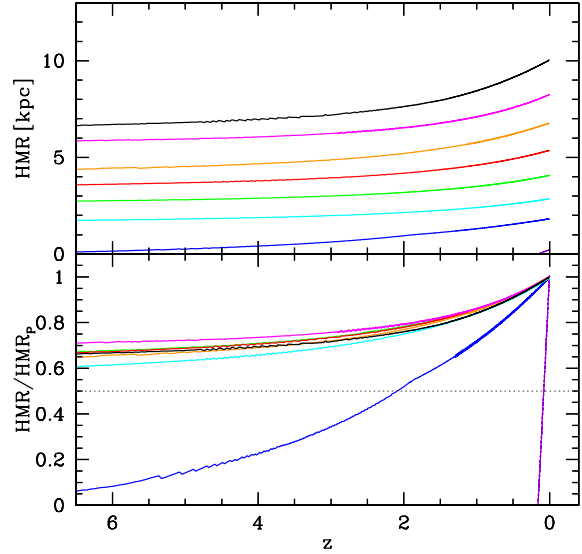}
\caption{a) The evolution of the radius enclosing the half baryonic
mass of disc, $HMR$, with redshift $z$. b) The redshift evolution
of the same half mass radii, normalised to their final value.  Colour
lines have the same coding as in Fig.~1.}
\label{reffm}
\end{figure}

Figure~\ref{reffm} shows the evolution of the half (baryonic) mass
radius of the disc, $HMR$, with redshift, for the same galaxies of
Fig.~\ref{masas}. This figure also supports the notion that the growth
of massive discs is more rapid than the growth of the low mass ones
(which are still forming their discs).
In panel a) we see directly the redshift evolution of these
HMRs. Clearly, the lowest mass galaxies evolve very late compared with
the others. This difference of behaviour is even clearer in panel b),
where the HMRs are normalised to their final value reached at the end
of the evolution (or the present time), $HMR_{P}$. In that case, most
of galaxies show a similar evolution, except the lowest mass galaxies,
which have very separated evolutionary tracks.

\begin{figure}
\includegraphics[width=0.35\textwidth,angle=-90]{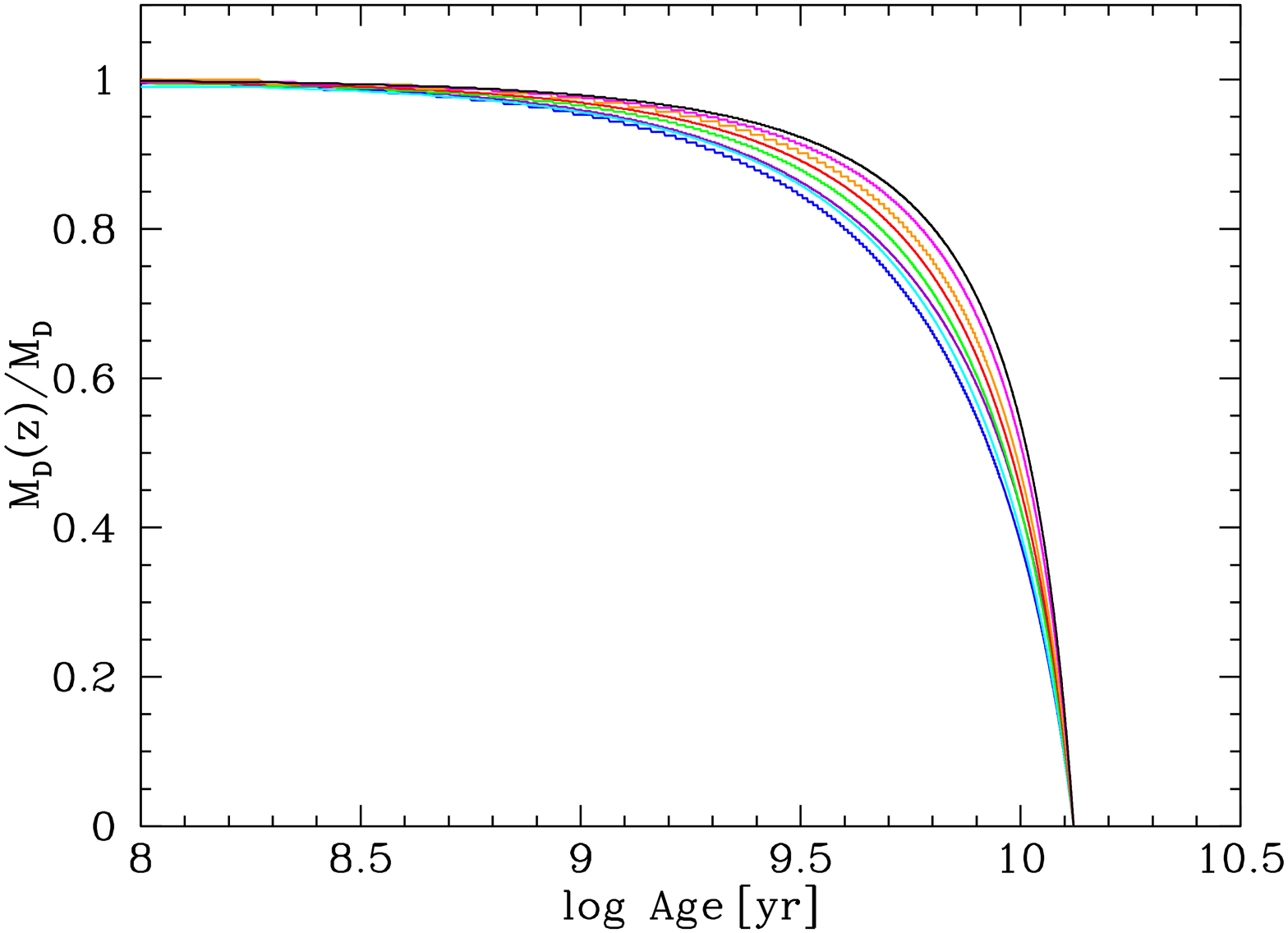}
\includegraphics[width=0.35\textwidth,angle=-90]{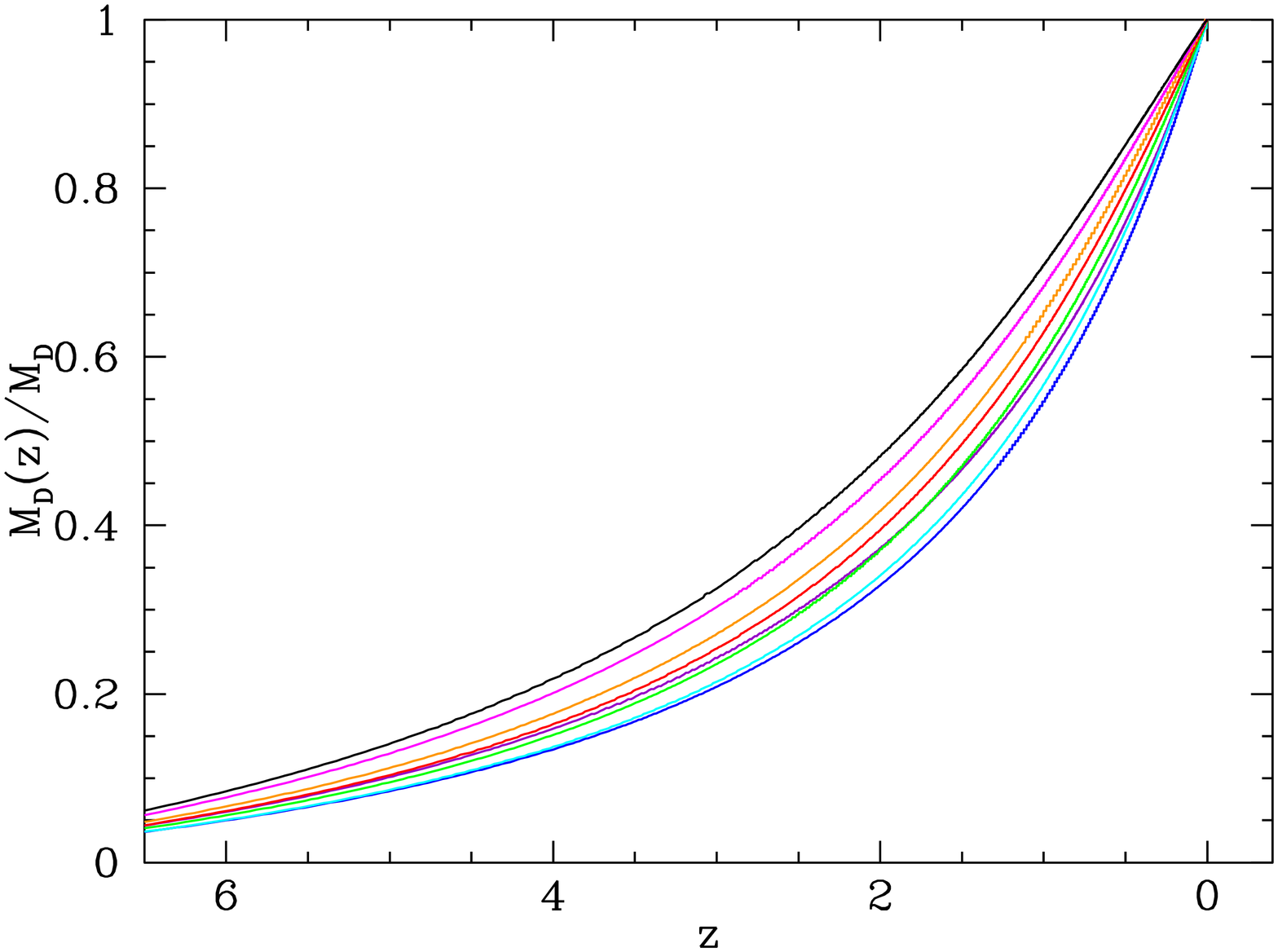}
\caption{The evolution of the masses of  discs normalised to their final values 
as a function: a) of the disc age in logarithmic scale; b) of the redshift.}
\label{md_z}
\end{figure}

We also check how the discs grow by comparing the fraction of the
final disc mass which is in place at each redshift in
Fig.~\ref{md_z}. Obviously, all tracks end at a value of unity at
$z=0$ (100\% of $M_{\textrm{\scriptsize D}}$ in the present time), but
the evolution differs for each galaxy depending on the virial mass.
This kind of plot is similar to that obtained by \citet{perez13}, for
the {\sc CALIFA} survey \citep[Calar Alto Legacy Integral Field
Area,][]{sanchez12} galaxies as shown in panel a), where we represent
the proportion of the total mass placed in the disc as a function of
the disc age, equal to $13.8-t$, $t$ being the evolutionary time. In
panel b) the same is shown as a function of the redshift $z$.

\begin{figure}
\includegraphics[width=0.35\textwidth,angle=-90]{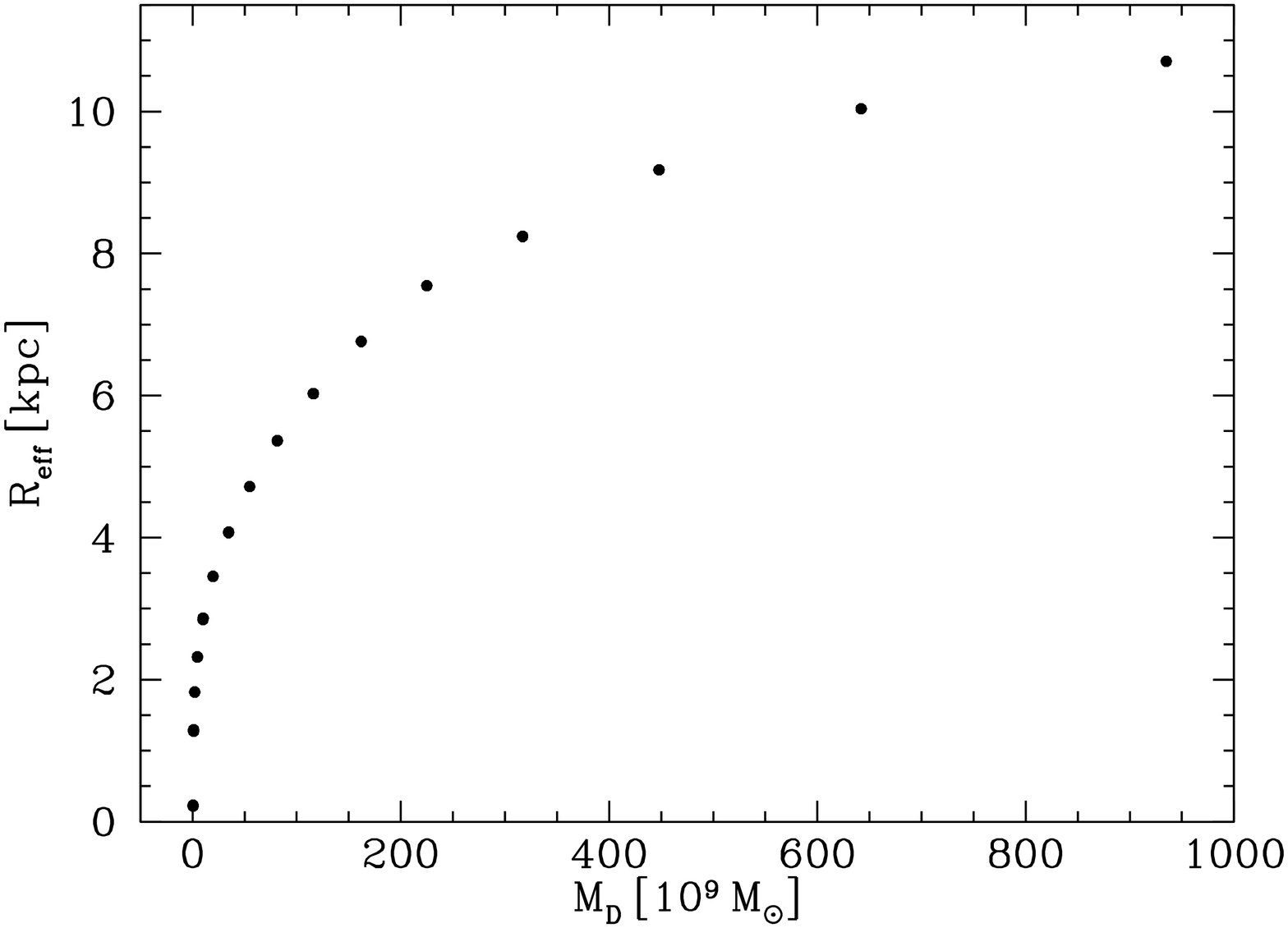}
\includegraphics[width=0.35\textwidth,angle=-90]{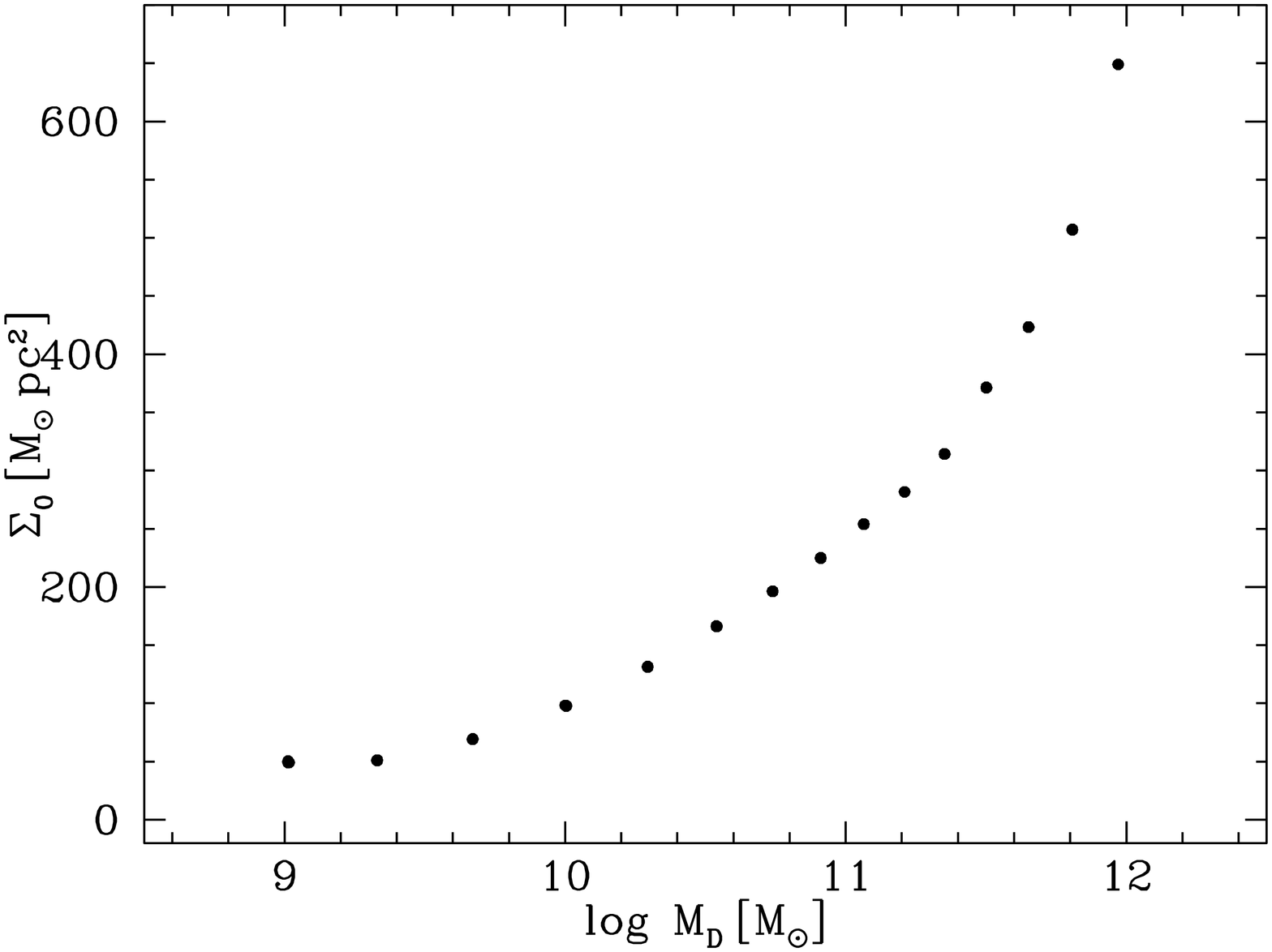}
\caption{a) The present half mass radius $HMR_{P}$  and 
b) The present central surface density $\Sigma_{0}$, as a function 
of the final mass of disc $M_{\textrm{\scriptsize D}}$, on a
logarithmic scale.}
\label{hmr_md}
\end{figure}

On the other hand we may also compare the final $HMR_{P}$ or radius at
the present time, for different galaxies. Panel a) of
Fig.~\ref{hmr_md} plots $HMR_{P}$ as a function of the disc mass,
$M_{\textrm{\scriptsize D}}$, also at the present time.  Clearly,
there is a relationship between the baryonic mass in discs and their
sizes, measured by the $HMR_{P}$, as expected. Actually, by assuming
that discs are exponential and using these quantities, we also might
calculate the surface density at the center of the galaxy as
$\Sigma_{0}=\frac{M_{\textrm{\scriptsize D}}}{4\pi R^{2}}$. We plot
this surface density, $\Sigma_{0}$, as a function of the disc mass in
panel b) of the same Fig.~\ref{hmr_md}, where points show a smooth
behaviour.

\section{Discussion}
\label{dis}

By imposing the the final disc mass distributions resemble those from
\citet{sal07}, we have computed the infall rates in the discs of
spiral galaxies.  From our models, we have reached important results
regarding the mass assembly in disc galaxies compared with previous
results in the literature.  The new prescriptions adopted in this work
resulted in a MWG analogue where radial regions with $R > 10$ kpc have
collapse timescales larger than the Hubble time (see
Fig. \ref{tcol_r}). Other radial functions for $\tau$ in the
literature have also suggested these timescales are reached at
comparable $\sim$10-15~kpc galactocentric radii; as such, a larger
portion of the outer disc in our new MWG analogue is still currently
assembling, compared to conclusions which would be drawn with most of
the previously available timescales in the literature.  In
Fig. \ref{infall-r}, a clear dependence of the gas infall rate on
redshift, and also on radius, appears except for the outer disc
regions where the infall rate is almost constant in spite of redshift
variations. At the inner disc regions and the bulge, however, the cold
gas accretion is higher at high redshifts than now.  This cold gas
accretion in the inner regions of high-redshift galaxies is supported
by observations: \citet{cresci10} find that a massive infall of gas in
the central regions of high-redshift galaxies ($z \sim 2 - 3$) is
needed to explain the gas metallicity distribution within these
galaxies. In the local universe there is also evidence that there
still exists cold gas accretion in disc galaxies \citep[e.g.][and
references therein]{jorge14b}.

With our collapse timescales varying with $R$, the resulting discs are
also created in agreement with the inside-out formation scenario.
Following Fig. \ref{infall-r}, the inner regions of a galaxy are
formed faster than the outer ones, while these latter present infall
rates which are less variable with redshift. In agreement with this,
the outer regions of the MWG analogue model are still accreting gas.
This new result is in agreement with surveys of extended UV discs
around both early- and late-type galaxies in the local universe that
suggest gas accretion in their outer parts \citep{lem11, mof12}. In
fact, these latter authors establish that gas-rich galaxies below a
stellar mass of $M_{*}\sim 5 \times 10^{9}$\,Msun\ display UV-bright
discs, supporting an active disc growth.  This limit is basically the
value in which $\tau_{c}> T_{UNIVERSE}$ in our models, as seen in
Table 1, between models 4 and 5. This limit in the collapse timescale
also appears when considering the different radial regions within
discs, thus suggesting that these discs may continue growing at the
present time even in the most massive galaxies.

Another important result that we have obtained is the dependence on
redshift of the infall rate normalised to the disc mass in each
time-step for different \Mvir, shown in Fig. \ref{infall-masa}. The
gas accretion profile is similar for all simulated galaxies at $z >
2$, and only for lower redshifts does the infall rate depend on
\Mvir. This has important implications for how discs grow their masses
at high redshifts. Since galaxies form and evolve by accreting gas
from their surroundings, our models suggest that the accretion of gas
is similar for spiral galaxies, regardless of their
masses. \citet{keres09} showed that cold mode accretion via filaments
is responsible for the supply of gas in
galaxies. \citet{kereshernquist09} showed that the gas is accreted via
filaments from the intergalactic medium and these filaments can
condense into clouds and provide gas for star formation. These clouds
may be analogs of HVCs surrounding the MWG which are thought to
provide the fuel for star formation in the Galaxy. Therefore, the
similarity between the accretion rates found here for spiral galaxies
with different \Mvir\ suggests an analogous disc growth for these
galaxies and, consequently, similar disc properties. We will analyse
these properties, as the possible formation of stars in the outer
regions of discs or the variations of radial gradients of abundances,
in a future work (Moll\'{a} et al., in preparation). Therein, we will
take into account the star formation processes and the consequent
metal enrichment, not included yet here, when the chemical evolution
computed is computed self-consistently.

\section{Conclusions}
\label{conclusions}
In this work we have presented the infall rates used as input to a
grid of chemical models for spiral galaxies.
We also performed a systematic comparison with data and
competing models in the literature, including cosmological
simulations.  Our main conclusions can be summarised as follows:

\begin{enumerate}
\item The more massive the galaxy, the higher the absolute value of
the infall rate at all redshifts.
\item The infall rates necessary to reproduce the relationship between
\Mvir$-M_{\textrm{\scriptsize D}}$ are smoother than the ones assumed
in classical chemical evolution models, including our earlier
generation of models (MD05).  They are also lower and smoother than
the accretion rates produced in cosmological simulations which create
spheroids and are in better agreement with those resulting in
realistic late-type discs (employing contemporary prescriptions for
star formation and feedback).
\item The evolution of the infall rate with redshift is quite constant
for discs, decreasing smoothly for $z < 2$ and showing very similar
behaviour for all radial regions, with differences only in the absolute
value. This smooth evolution is different than the steep decline with
redshift shown for bulges.
\item The normalised infall rate is essentially the same for all discs 
until $z=2$, and shows only small differences for $z< 2$.
\item The final relationship among disc and halo masses, 
denoted as SHMR, obtained by these new prescriptions, agrees
well with observations and cosmological simulations.
\item From redshift $z=2.5$ to today, discs grow in size by a factor of 
two (except for the lowest mass galaxies), 
while the disc masses increase by a factor of five to ten. 
\item These infall rates (decreasing with time and with radius within
each theoretical galaxy) are in agreement with the classical
inside-out scenario, and also with the cosmological simulations infall
rates from \citet{courty10}.
\item The growth of discs continue to the present time, with
gas accretion across the discs of low mass systems, and in the outer
regions of more massive spirals.
\end{enumerate}

\section{Acknowledgments}
We gratefully acknowledge the advice of the referee which improved the
manuscript dramatically.  This work has been supported by
DGICYT grant AYA2010-21887-C04-02 and AYA2013-47742-C4-4-P.  MM thanks
the kind hospitality and wonderful welcome of the Jeremiah Horrocks
Institute at the University of Central Lancashire, the E.A. Milne
Centre for Astrophysics at the University of Hull, and the
Instituto de Astronomia, Geof\'{\i}sica e Ci\^{e}ncias
Atmosf\'{e}ricas in S\~{a}o Paulo (Brazil), where this work was
partially done.  This work has been supported financially by grant
2012/22236-3 from the S\~{a}o Paulo Research Foundation (FAPESP). This
work has made use of the computing facilities of the Laboratory of
Astroinformatics (IAG/USP, NAT/Unicsul), whose purchase was made
possible by the Brazilian agency FAPESP (grant 2009/54006-4) and the
INCT-A.  We acknowledge PRACE, through its Distributed Extreme
Computing Initiative, for resource allocations on Sisu (CSC, Finland),
Archer (EPCC, UK), and Beskow (KTH, Sweden). We acknowledge the
support of STFC DiRAC High Performance Computing Facilities; DiRAC is
part of the UK National E-infrastructure.

\label{lastpage}
\end{document}